\documentclass[12pt]{iopart}
\usepackage{bm,graphics,graphicx,epsfig,color}
\usepackage{iopams}
\def\be{\begin{equation}}
\def\ee{\end{equation}}

\def\gsim{\mathrel{
\rlap{\raise 0.511ex \hbox{$>$}}{\lower 0.511ex
\hbox{$\sim$}}}}
\def\lsim{\mathrel{
\rlap{\raise 0.511ex \hbox{$<$}}{\lower 0.511ex
\hbox{$\sim$}}}}
\begin{document}
\title{ 
LISA as a dark energy probe}

\author{K.G. Arun$^1$$^{,2}$$^{,6}$, Chandra Kant Mishra$^3$$^{,5}$, Chris Van Den Broeck$^4$, B.R. Iyer$^3$, B.S. Sathyaprakash$^4$, Siddhartha Sinha$^3$$^{,5}$}
\ead{arun@physics.wustl.edu, chandra@rri.res.in, Chris-van-den-Broeck@astro.cf.ac.uk, bri@rri.res.in, B.Sathyaprakash@astro.cf.ac.uk, p siddhartha@rri.res.in}
\address{$^1$Institut d'Astrophysique de Paris, UMR 7095--CNRS, Universit\'e Pierre et Marie Curie, 98$^{bis}$ Boulevard Arago, 75014 Paris, France\\
$^2$LAL, Universit{\'e} Paris-Sud, IN2P3/CNRS, Orsay, France\\
$^3$Raman Research Institute, Sadashivanagar, 
Bangalore 560080, India\\
$^4$School of Physics and Astronomy, Cardiff University, Queen's Buildings, The Parade, Cardiff CF24 3YB, UK\\ 
$^5$Department of Physics, Indian Institute of Science, Bangalore, 560 012, India\\
$^6$ Presently at: McDonnell Centre for Space Sciences, Department of Physics, Washington University, St Louis, MO 63130, USA
}

\begin{abstract}

Recently it was shown that the inclusion of higher signal harmonics in the inspiral signals of binary supermassive black holes (SMBH) leads to dramatic improvements in parameter estimation with the Laser Interferometer Space Antenna (LISA). In particular, the angular resolution becomes good enough to identify the host galaxy or galaxy cluster, in which case the redshift can be determined by electromagnetic means. The gravitational wave signal also provides the luminosity distance with high accuracy, and the relationship between this and the redshift depends sensitively on the cosmological parameters, such as the equation-of-state parameter $w=p_{\rm DE}/\rho_{\rm DE}$ of dark energy. Using binary SMBH events at $z < 1$ with appropriate masses and orientations, one would be able to constrain $w$ to within a few percent. We show that, if the measured sky location is folded into the error analysis, the uncertainty on $w$ goes down by an additional factor of 2-3, leaving weak lensing as the only limiting factor in using LISA as a dark energy probe.

\end{abstract}

\date{\today}
\pacs{04.30.Db, 04.25.Nx, 04.80.Nn, 95.55.Ym}
%\maketitle

\section{Introduction}

An important problem in present-day cosmology is the nature of dark energy (for a review, see, e.g., \cite{PeeblesRatra03}). Assuming a homogeneous and isotropic Universe, dark energy can be characterized by a parameter $w(z) = p_{\rm DE}(z)/\rho_{\rm DE}(z)$, where $p_{\rm DE}$ and $\rho_{\rm DE}$ are the pressure and density, respectively. A constant value of $w = -1$ would correspond to a positive cosmological constant in the Einstein equations, but other possibilities are by no means excluded by current constraints on $w$. From the five year WMAP data combined with supernovae measurements as well as baryon acoustic oscillations in the galaxy distribution, one has $-1.11 < w < -0.86$ at the 95\% confidence level \cite{Hinshaw08}. 

As argued in \cite{HolzHugh05,DaHHJ06}, by treating binary supermassive black holes (SMBH) as ``standard sirens", LISA could play an important role in investigating the physical origin of dark energy. However, these works assumed the so-called restricted post-Newtonian waveform (RWF) for quasi-circular, adiabatic inspiral, which only contains the dominant harmonic at twice the orbital frequency and no corrections to the amplitude. The full waveform (FWF) also includes other harmonics, each having post-Newtonian (PN) corrections to their amplitudes, which, when taken into account, have a significant effect on parameter estimation, as shown in \cite{SinVeccHellingsMoore,ChrisAnand06b} in the context of both LISA and ground-based detectors. More recently, the implications of adding the merger and ringdown phases of the binary
SMBH coalescence for angular resolution of LISA were examined \cite{BHHS08}. 

The estimation of the luminosity distance and sky location with LISA was recently studied with inspiral waveforms at high order in amplitude \cite{AISSV,SintesAndTrias,PorterCornish08}, and the improvements were found to be dramatic. In particular, unlike what has been suggested based on the results of using RWF, for a wide range of systems LISA will have sufficient angular resolution to identify the host galaxy or galaxy cluster. Following an idea by Schutz \cite{Schutz86}, one could then use SMBH coalescences to constrain cosmological parameters. Knowing the host one can measure the redshift at which the inspiral event occured, and the gravitational-wave signal itself gives the luminosity distance with high precision. The relationship between the two depends very sensitively on the Hubble constant, $H_0$, the normalized matter and dark energy densities, $\Omega_{\rm M}$ and $\Omega_{\rm DE}$, and the dark energy equation-of-state parameter $w$. Assuming that the first three are known with high accuracy (and indeed, $w$ is currently the worst constrained \cite{Hinshaw08}), it was found that binary SMBH merger events at $z < 1$ could allow us to constrain $w$ to within a few percent, if $w$ is constant in time \cite{AISSV}. (Possible time dependence of $w$ will also have to be investigated in the future.) Event rates within that redshift range are uncertain, although one of the SMBH formation models predicts a rate as large as $\sim 10\,\mbox{yr}^{-1}$ (see \cite{Nelemans06} for an overview). LISA may be able to constrain $w$ at the same level as dedicated dark energy probes, but foregoing the lower rungs of the cosmic distance ladder.

\emph{A priori}, a signal would be matched-filtered against a bank of templates in which the two sky position angles are free variables. Due to LISA's noise, the templates that match best will have position parameters that differ slightly from the true values. Using, e.g., the Fisher matrix formalism, one can arrive at an error box in the sky containing the host cluster of the merger event. If the host can be identified by electromagnetic means, the sky position will be known with negligible error. One would then revisit the LISA data and try to match the signal using only template waveforms that correspond to the measured sky location. The error analysis would also be redone, this time excluding the sky position variables, thus reducing the dimensionality of the parameter space. Since in the gravitational-wave signal, the luminosity distance $D_{\rm L}$ is strongly correlated with sky position, the error on $D_{\rm L}$ will be further reduced, leading to an even tighter constraint on $w$. This will be the focus of the present study. As we will show, in those cases where identification of the host cluster is possible, weak lensing will be the only limiting factor in the determination of $w$.

\section{Waveform models, noise curve, and parameter estimation}
  
To date, the highest-order waveforms available are 3.0PN in amplitude and 3.5PN in phase \cite{3PN,BFIJ02BDEI04}, and these are the ones we will use here. For a noise curve we take that of the latest Mock LISA Data Challenge (MLDC) \cite{MLDC}. We will study errors for lower cut-off frequencies $f_{\rm s} = 10^{-4}$ Hz and $f_{\rm s} = 10^{-5}$ Hz. 

Since the systems we will consider only have power at frequencies $f \lesssim 5 \times 10^{-3}$ Hz, it is reasonable to model LISA as a combination of two interferometers having two arms each; these are usually referred to as detectors I and II. We use the stationary phase approximation (SPA) to the Fourier transform of the waveform. (As a minor caveat, we note that windowing effects start affecting the accuracy of SPA at 3.5PN in amplitude \cite{DrozetalDIS}.) The full signal up to 3.0PN in amplitude as seen in each of the two detectors is a superposition of eight harmonics of the orbital frequency, taking the general form \cite{AISS07} 
\begin{eqnarray}
\tilde{h}_{\rm I, II}(f) &=& \frac{\sqrt{3}}{2} \frac{2M\nu}{D_L}
\,\sum_{k=1}^{8}\,\sum_{n=0}^6\,
\frac{A^{\rm I, II}_{(k,n/2)}(t(f_k))
\,x^{\frac{n}{2}+1}(t(f_k))\,e^{-i\phi^{\rm I, II}_{(k,n/2)}(t(f_k))}}{2\sqrt{k\dot{F}(t(f_k))}}  \nonumber\\
&&\,\,\,\,\,\,\,\,\,\,\,\,\,\,\,\,\,\,\,\,\,\,\,\,\,\,\,\,\,\,\,\,\,\,\,\,\,\,\,\,\,\,\,\,\,\,\,\,\,\,\,\,\,\times\,\exp\left[i\,\psi_{f,k}(t(f_k))\right],
\label {FT}
\end{eqnarray}
where $f_k\equiv f/k,$ an overdot denotes derivative with respect to time, and
$\psi_{f,k}(t(f_k))$ is given by
\begin {equation}
\psi_{f,k}(t(f_k)) = 2\pi f\,t(f_k) -
k\,\Psi(t(f_k))-k\,\phi_{\rm D}(t(f_k)) - \pi/4.
\label {phase}
\end {equation}
Quantities in Eqs.\ (\ref{FT}) and (\ref{phase}) with the argument $t(f_k)$
denote their values at the time when the instantaneous orbital frequency $F(t)$ sweeps
past the value $f/k$ and $x(t)$ is the PN parameter given by
$x(t) = (2\pi M F(t))^{2/3}$. $A^{\rm I, II}_{(k,n/2)}(t)$ and $\phi^{\rm I, II}_{(k,n/2)}(t)$ are the polarization
amplitudes and phases of the $k$th harmonic at $n/2$th PN order in amplitude. 
$\Psi(t)$ is the orbital phase of the binary and $\phi_{\rm D}(t)$ is a time-dependent 
term representing Doppler modulation. For more details on all of these quantities, see \cite{AISSV} and references therein. 

The \emph{restricted} waveform (RWF) corresponds to retaining the term with 
$k=2$ and $n=0$ in Eq.~(\ref{FT}) and neglecting all others. The RWF 
has the dominant harmonic at twice the orbital 
frequency but no other harmonics, nor PN corrections to the amplitude of the dominant
one. It does, however, include the post-Newtonian expansion of the phase
to all known orders, i.e., up to 3.5PN. 

Each harmonic in $\tilde h_{\rm I, II}(f)$ is taken to be zero outside a certain frequency range. The upper cut-off frequencies are dictated by the last stable orbit (LSO), beyond which the PN approximation breaks down. For simplicity we assume that this occurs when the orbital frequency $F(t)$ reaches $F_{\rm LSO} = 1/(6^{3/2} 2\pi M)$ -- the orbital frequency at LSO of a test particle in Schwarzschild
geometry in $c=G=1$ units. Consequently, in the frequency domain, the contribution to $\tilde{h}_{\rm I,II}(f)$ from the $k$th harmonic is set to zero for frequencies above $k F_{\rm LSO}$. In determining the lower cut-off frequencies we assume that the source is observed for at most one year, and the $k$th harmonic is truncated below a frequency $k F_{\rm in}$, where $F_{\rm in}$ is the value of the orbital frequency one year before LSO is reached \cite{AISS07}. Thus, we take the lower cut-off frequency of the $k$th harmonic to be the maximum of $k F_{\rm in}$ and the detector's lower cut-off frequency $f_{\rm s}$.   

Following earlier work \cite{Cutler98,Hughes02,BBW05a,ALISA06,BarackCutler04}
we employ the Fisher matrix approach \cite{FinnChernoff} to the problem of parameter
estimation. The waveforms depend on nine parameters which are chosen to be
\begin{equation}
{\mathbf p} \equiv \left(\ln{\cal M},\delta,t_{\rm C},\phi_{\rm C},\mu_{\rm L},\phi_{\rm L},\mu_{\rm S},\phi_{\rm S},\ln D_{\rm L}\right),
\label{eq:params}
\end{equation}
where $\delta\equiv (m_2-m_1)/M$, with $m_1 \leq m_2$ the individual component masses. 
$t_{\rm C}$ and $\phi_{\rm C}$ are, respectively, the time and orbital phase at coalescence. Below we will consistently set the \emph{values} of $t_{\rm C}$ and $\phi_{\rm C}$ to zero, but both parameters are included as coordinates on the space of signals in computing the Fisher matrix. 
$\mu_{\rm S} = \cos \theta_{\rm S}$ and $\phi_{\rm S}$ determine the source position in the 
sky while $\mu_{\rm L} = \cos \theta_{\rm L}$ and $\phi_{\rm L}$ determine the orientation
of the binary's orbit with respect to a nonrotating detector at the solar system
barycenter. Following Ref.~\cite{Cutler98}, we have fixed the initial position and
orientation of LISA by setting the constants $\phi_0$ and $\alpha_0$ defined there to zero at the time of coalescence.

Note that our waveforms do not include spin, which would be somewhat awkward when using the stationary phase approximation. In principle we could have assumed static spins as in \cite{SintesAndTrias}. This can lead to much larger uncertainties than in the non-spinning case \cite{PoissonWill}, yet at least for RWF, genuinely \emph{dynamical} spins are known to greatly improve parameter estimation \cite{Vecchio04,LangHughes06}. We expect further improvement when combining harmonics and spin, albeit not necessarily a large one, since the two tend to break the same degeneracies. In any case, if spins are left out of the Fisher matrix altogether, errors are likely to be \emph{closer} to the ``true" ones than when static spins are included. 

As usual, the Fisher matrix ${\mathbf \Gamma}$ for LISA as a whole is simply ${\mathbf \Gamma} = {\mathbf \Gamma}_{\rm I} + {\mathbf \Gamma}_{\rm II}$, where ${\mathbf \Gamma}_{\rm I, II}$ are the Fisher matrices computed from the waveforms $\tilde{h}_{\rm I, II}(f)$. The parameters used will be the ones listed in Eq.~(\ref{eq:params}), so that \emph{a priori}, ${\mathbf \Gamma}$ is a $9\,\times\,9$ matrix. The errors in the 
estimation of $\mu_{\rm S}$ and $\phi_{\rm S}$ obtained in this way will be converted to 
a solid angle $\Delta \Omega_{\rm S}$ centered around the actual source direction.
Following the notation of \cite{BarackCutler04},
\begin{equation}
\Delta \Omega_{\rm S}=  2\pi \sqrt{(\Delta \mu_{\rm S} \,\Delta
\phi_{\rm S})^2-\langle\delta \mu_{\rm S}\,\delta \phi_{\rm S}\rangle^2},
\end{equation}
where the second term is the covariance between $\mu_{\rm S}$ and  
$\phi_{\rm S}$. 

If the sky location of the host galaxy or cluster is known, then one can redo the error analysis with a $7\,\times\,7$ Fisher matrix ${\mathbf \Gamma'}$ in which the variables $(\mu_{\rm S},\phi_{\rm S})$ are not taken into account. The construction of this smaller Fisher matrix is completely analogous to that of ${\mathbf \Gamma}$. 

In what follows, whenever it is necessary to consider a specific cosmological model we will assume a spatially flat, homogeneous and isotropic Universe with the Hubble constant $H_0 = 75\,\mbox{km}\,\mbox{s}^{-1}\mbox{Mpc}^{-1}$, dark energy equation-of-state parameter $w=-1$, matter density $\Omega_{\rm M}=0.27$, and dark energy density $\Omega_{\rm DE}=0.73$, with $\Omega_{\rm Total}=\Omega_{\rm M} + \Omega_{\rm DE} = 1$. Below we will refer to this as our ``fiducial" model.

\section{Determination of the dark energy equation-of-state parameter}

The relationship $D_{\rm L}(z)$ between luminosity distance and redshift depends very sensitively on the values of $H_0$, $\Omega_{\rm M}$, $\Omega_{\rm DM}$, and $w$. Assuming a flat FLRW Universe and a time-independent $w$ (as we shall do throughout this paper),  
\begin{equation}
D_{\rm L}(z) = (1+z)\,\int_0^z \frac{dz'}{H_0 \left[ \Omega_{\rm M}(1+z')^3 + \Omega_{\rm DE} (1+z')^{3(1+w)} \right]^{1/2}}.
\label{DLgeneral}
\end{equation}
Suppose a gravitational wave signal from a binary SMBH inspiral event is detected. Then from the signal itself, $D_{\rm L}$ can be determined, but not the redshift $z$. However, because of both amplitude and phase modulation in the waveform due to LISA's motion around the Sun and the time dependence of its orientation, it will be possible to obtain a box in the sky which contains the host. If the number of galaxy clusters within this error box is not too large then the host cluster of the event can be identified, and the redshift can be obtained with negligible error by conventional electromagnetic means\footnote{Clusters may not be well-defined at red-shifts of interest to us; we use the number of clusters only as a quantitative way of judging if the sky resolution is good enough for identifying the host.}. 

Given a binary SMBH merger event, one can estimate how many galaxy clusters within the sky error box $\Delta \Omega_{\rm S}$ need to be taken into account as potential hosts. Note that we can \emph{not} use the volume error box fixed by both $\Delta\Omega_{\rm S}$ and $\Delta D_{\rm L}$: We want to estimate $w$ through the relationship between luminosity distance and redshift, so that the two need to be obtained independently. We will proceed as follows. Using the ``fiducial" cosmological model described at the end of the previous section, we associate a fiducial redshift value $z_0$ with the value of $D_{\rm L}$ measured from the inspiral signal. The comoving volume inside the cone spanned by $\Delta\Omega_{\rm S}$ and stretching up to the redshift $z_0$ is given by:
\be
V_{\rm C} = \int_0^{z_0} dz' \frac{\Delta\Omega_{\rm S}}{H_0}\frac{D^2_{\rm L}(z')}{(1+z')^2}\frac{1}{\sqrt{\Omega_{\rm M}(1+z')^3 + \Omega_{\rm DE}(1+z')^{3(1+w)}}}.
\ee 
Multiplying $V_{\rm C}$ by the number density of clusters we obtain an estimate for ${\mathbf N_{\rm clusters}}$, the number of clusters that need to be considered as potential hosts of the inspiral event. At high redshifts the cluster density is not known very well; following Ref.~\cite{Bahcall:2003} we will take it to be $\sim\,2 \times 10^{-5} h^3\mbox{Mpc}^{-3}$, where $h$ is the value of the Hubble constant at the current era in units of $100\,\mbox{km}\,\mbox{s}^{-1}\,\mbox{Mpc}^{-1}$ (i.e., $h = 0.75$ for our fiducial cosmological model).

At this point one might object that we should also take into account clusters with a much higher redshift than our fiducial $z_0$; after all, redshift is precisely what we want to measure. As it turns out, this is not necessary to obtain a reasonable estimate of ${\mathbf N_{\rm clusters}}$. As an example, consider a system at distance $D_{\rm L} = 3$ Gpc, which in our fiducial cosmological model corresponds to $z_0 = 0.55$. Suppose one wanted to consider a potential host cluster at $z=0.6$. Then in order to reconcile this slightly larger redshift with the measured $D_{\rm L}$, for the same values of $H_0$, $\Omega_{\rm M}$, and $\Omega_{\rm DE}$ one would have to assume $w = -0.47$, a value that is strongly excluded by WMAP and supernovae observations \cite{Hinshaw08}. Uncertainties in the other cosmological parameters would also have to be taken into account, but in practice it would probably not be necessary to consider potential hosts at redshifts that differ from the fiducial $z_0$ by more than 20 percent. Our aim is to get a rough estimate for ${\mathbf N_{\rm clusters}}$, and for that purpose, the method outlined above will suffice.

If ${\mathbf N_{\rm clusters}}$ comes out to be of order 1 then the host cluster can be identified and a redshift value can be obtained. Finding the host may well be possible even if ${\mathbf N_{\rm clusters}} \gg 1$, since the SMBH merger event could be accompanied by a distinctive electromagnetic counterpart, which might be found using large survey instruments \cite{LangHughes08,Kocsisetal08}. Nevertheless, in this paper we (arbitrarily) choose ${\mathbf N_{\rm clusters}} < 3$ as a localizability criterion.  

We now turn to measuring $w$. If the host can be localized then the redshift $z$ can be determined with negligible error, and values for the cosmological parameters can be obtained from $D_{\rm L}$ and $z$ through the relation (\ref{DLgeneral}). In practice this will require multiple measurements, since what is measured directly is only $D_{\rm L}$. In a complete analysis one would have to estimate uncertainties on the four parameters $w$, $H_0$, $\Omega_{\rm M}$, and $\Omega_{\rm DE}$, and their correlations, all at once from the LISA data itself, with input from gravitational-wave observations of other LISA sources such as extreme mass ratio inspirals (EMRIs) \cite{Hogan}, or observations with ground-based gravitational-wave observatories like Advanced LIGO or Einstein Telescope \cite{ET}. With Einstein Telescope it should be possible to determine the four unknowns ($w$, $H_0$, $\Omega_{\rm M}$, and $\Omega_{\rm DE}$) by fitting $D_{\rm L}(z)$ to observed data using large numbers of stellar mass inspiral events (which give $D_{\rm L}$) with electromagnetic counterparts (which give $z$). The number of observable inspirals might be as large as $500\,\mbox{yr}^{-1}$, giving several thousands over a period of five years. We are , currently evaluating the covariance matrix in the cosmological parameters associated with such observations\footnote{In principle we could also consider combining results from gravitational-wave observations with those of expected conventional probes, but the great benefit of gravitational-wave astronomy for cosmology will precisely be that cosmological parameters would be measurable while foregoing the lower rungs of the cosmic distance ladder.}. 

In the present study we are only interested in getting a rough sense of the level of accuracy we can expect in extracting $w$; consequently we neglect the uncertainties on $H_0$, $\Omega_{\rm M}$, and $\Omega_{\rm DE}$, and we assume that $w$ does not depend on time. The error on $w$ can then be estimated as: 
\be
\Delta w = D_{\rm L} \left| \frac{\partial D_{\rm L}}{\partial w} \right|^{-1}\frac{\Delta D_{\rm L}}{D_{\rm L}}.
\label{Deltaw}
\ee

Since \emph{a priori} one must assume that sky position is unknown, first,  all of the parameters in Eq.~(\ref{eq:params}) need to be included in the error analysis. Indeed, before the host cluster has been identified, the signal must be matched against a family of template waveforms in which the sky position parameters $(\mu_{\rm S}, \phi_{\rm S})$ are free variables. However, once the host has been identified electromagnetically, the sky position will be known with negligible error. One can then revisit the gravitational wave data and match the signal against a smaller template family in which $(\mu_{\rm S},\phi_{\rm S})$ are set to the measured values. The set of unknown parameters now comprises only 
\begin{equation}
{\mathbf p'} \equiv \left(\ln{\cal M},\delta,t_{\rm C},\phi_{\rm C},\mu_{\rm L},\phi_{\rm L},\ln D_{\rm L}\right).
\label{eq:paramsprime}
\end{equation}
The errors on these can be estimated by constructing a $7\,\times\,7$ Fisher matrix ${\mathbf \Gamma'}$ in the usual way. But because luminosity distance is strongly correlated with sky position (see Appendix B of Ref.~\cite{AISSV}), we may expect the error $\Delta D_{\rm L}/D_{\rm L}$ computed from ${\mathbf \Gamma'}$ to be significantly smaller than the one obtained from the original Fisher matrix ${\mathbf \Gamma}$. Since $\Delta D_{\rm L}/D_{\rm L}$ determines the error on $w$ though Eq.~(\ref{Deltaw}), $\Delta w$ will be proportionally smaller. In Ref.~\cite{AISSV}, only the full Fisher matrix ${\mathbf \Gamma}$ was employed to estimate $\Delta D_{\rm L}/D_{\rm L}$ even in cases where it was possible to identify the host cluster, so that the values for $\Delta w$ given there are overestimates. Here we will take the extra step of recomputing $\Delta w$ from the smaller Fisher matrix ${\mathbf \Gamma'}$.

\section{Results}

\begin{table}
	\centering
		\begin{tabular}[t]{c|c|c|c|c|c|c|c|c|c}
		\hline
		\hline
      $\mu_{\rm S}$ & $\varphi_{\rm S}$  & $\mu_{\rm L}$ & $\varphi_{\rm L}$ & Model &   SNR  & $\Delta \ln D_{\rm L}$ 
     & $\Delta \Omega_{\rm S}$  & $\mathbf{N_{\rm clusters}}$ & $\Delta w$ \\ 
    & rad & & rad   &   &  & ($10^{-3}$)   &   ($10^{-6}$ sterad)  &  &  \\     
     \hline
     \multicolumn{10}{l}{{$(m_1,m_2) = (10^6,10^7)M_\odot $};\,\,\,{$f_{\rm s} = 10^{-4}$ Hz}} \\    
\hline
                     0.3    & 5 & 0.8   & 2 & RWF (before)  & 480   & 130    & 790    & \bf 16   & -- \\
                            &   &       &   & RWF (after)   & 480   &        &        & \bf      & --   \\
                            &   &       &   & 3.0PN (before)& 422   & 21     & 16     & \bf 0.32 & 0.12 \\
                            &   &       &   & 3.0PN (after) & 422   & 9.7    &        & \bf      & 0.055  \\
\hline
                     $-0.1$ & 2 &$-0.2$ & 4 & RWF (before)  & 749   & 120    & 8600   & \bf 170   & -- \\
                            &   &       &   & RWF (after)   & 749   &        &        & \bf       & -- \\
                            &   &       &   & 3.0PN (before)& 643   & 12     & 38     & \bf 0.78  & 0.068 \\
                            &   &       &   & 3.0PN (after) & 643   & 3.1    &        & \bf       & 0.017 \\
\hline
                     $-0.8$ & 1 & 0.5   & 3 & RWF (before)  & 1771  & 37     & 33000   & \bf 660  & -- \\
                            &   &       &   & RWF (after)   & 1771  &        &         & \bf      & --\\
                            &   &       &   & 3.0PN (before)& 1436  & 3.1    & 89      & \bf 1.8  & 0.017 \\
                            &   &       &   & 3.0PN (after) & 1436  & 1.2    &         & \bf      & 0.0068 \\  
\hline
                     $-0.5$ & 3 &$-0.6$ &$-2$& RWF (before) & 1212  & 80     & 20000  & \bf 410  & -- \\
                            &   &       &    & RWF (after)  & 1212  &        &        & \bf      & -- \\
                            &   &       &   & 3.0PN (before)& 1007  & 3.3    & 39     & \bf 0.80 & 0.019 \\ 
                            &   &       &   & 3.0PN (after) & 1007  & 1.8    &        & \bf      & 0.010 \\
\hline
                     0.9    & 2 & $-0.8$& 5 & RWF (before)  & 2419  & 1200   & 11000   & \bf 220  & -- \\
                            &   &       &   & RWF (after)   & 2419  &        &         & \bf      & --  \\
                            &   &       &   & 3.0PN (before)& 1781  & 1.9    & 22      & \bf 0.45 & 0.011\\
                            &   &       &   & 3.0PN (after) & 1781  & 0.69   &         & \bf      & 0.0039   \\
\hline
                     $-0.6$ & 1 &  0.2  & 3 & RWF (before)  & 1423  & 54   & 54000    & \bf 1100 & -- \\
                            &   &       &   & RWF (after)   & 1423  &      &          & \bf      & -- \\
                            &   &       &   & 3.0PN (before)& 1188  & 4.2  & 240      & \bf 4.9  & -- \\
                            &   &       &   & 3.0PN (after) & 1188  &      &          & \bf      & -- \\
\hline
                     $-0.1$ & 3 & $-0.9$& 6 & RWF (before)  & 1436  & 250    & 220000 & \bf 4600 & -- \\ 
                            &   &       &   & RWF (after)   & 1436  &        &        & \bf      & -- \\ 
                            &   &       &   & 3.0PN (before & 1215  & 12     & 3000   & \bf 63   & -- \\   
                            &   &       &   & 3.0PN (after) & 1215  &        &        & \bf      & -- \\ 
\hline
\hline
\end{tabular}
	\caption{Comparison of accuracy in LISA's measurement of various parameters for a lower cut-off frequency of $10^{-4}$ Hz, at 0PN and 3.0PN in amplitude for a binary with intrinsic masses $(10^6,10^7)M_\odot$, for different sky positions and orientations of the orbital plane. The distance is 3 Gpc, corresponding to a redshift $z_0=0.55$ in our fiducial cosmological model. Errors are shown both before and after taking into account the knowledge of the position of the host cluster. Whenever a dash appears in the $\Delta w$ column, it means that we deem the number of clusters in the sky error box to be too large for host identification; in that case a redshift can not be obtained and $w$ is unmeasurable.}
\label{t:1e61e7}
\end{table}

Let us consider an example. Table \ref{t:1e61e7} shows parameter estimation results for the restricted and the full waveforms at 3.0PN in amplitude\footnote{It is important to note that the MLDC noise curve used in this paper is different from (and more conservative than) the one used in \cite{AISSV}.}, comparing accuracies before and after a knowledge of the sky position has been taken into account, for a lower cut-off frequency of $10^{-4}$ Hz. The physical component masses are $(m_1,m_2) = (10^6,10^7)M_\odot$, at a luminosity distance $D_{\rm L} = 3$ Gpc (i.e., the observed masses are $(1+z_0)(10^6,10^7)M_\odot$ where $z_0 = 0.55$ in our fiducial model). Different choices are made for sky position and orientation of the plane of the inspiral. The focus is on ${\mathbf N_{\rm clusters}}$, which tells us how easy or difficult it would be to localize the host, and on $\Delta w$. Both for RWF and FWF, the 1-sigma uncertainties in parameters are compared before and after knowledge of the sky position is folded in. This affects $\Delta w$ through the value of $\Delta D_{\rm L}/D_{\rm L}$. 

What will be immediately apparent is the significant difference between errors from the restricted waveform and the full waveform. For the chosen system and lower cut-off frequency, with the RWF one cannot localize the host for any of the cases shown. The large improvements in going from RWF to FWF are as expected from the 2.5PN results of \cite{AISSV} as well as the 2.0PN simulations in \cite{SintesAndTrias}. The smaller sky error box in the case of the full waveform leads to smaller values of ${\mathbf N_{\rm clusters}}$ and makes it easier to localize the source; the equally dramatic improvement in distance estimation makes for a smaller error in $w$. 

What is new here is the dependence of errors on whether or not the information about the position of the source has been taken into account. As shown in \cite{AISSV} (see Appendix B of that paper), sky position is strongly correlated with luminosity distance. And indeed, we see improvements in $\Delta D_{\rm L}/D_{\rm L}$ -- and hence $\Delta w$ -- by factors of 2-3. We have also considered different masses from the ones in Table \ref{t:1e61e7}, such as $5\times (10^5,10^6)\,M_\odot$ and $3\times (10^6,10^7)\,M_\odot$, and found this trend to persist.

Our analysis does not take into account the effects of weak gravitational lensing by the matter distribution between the source and LISA. Weak lensing will distort the waveform and induce a systematic error in the estimation of $D_{\rm L}$, which for the distances we are considering will be at the level of 3-5 percent \cite{Kocsisetal2006}. On the other hand, it may be possible to partially remove the effect of weak lensing by mapping the mass distribution along the line of sight \cite{Gunnarssonetal2006}. In any case, our results indicate that whenever the host can be identified so that $w$ can be measured, weak lensing will essentially be the only limiting factor in the estimation of $w$, not the performance of LISA itself.

Depending on the time over which test masses in LISA can be kept in free fall, it may be possible to have a lower cut-off frequency of $10^{-5}$ Hz. The improvement in estimation of $\Delta\Omega_{\rm S}$ would be quite substantial; for the first choice of angles in Table \ref{t:1e61e7}, one would be able to localize the source even with the RWF. However, in those cases where the host could be identified also with $f_{\rm s} = 10^{-4}$ Hz, the improvement in $\Delta D_{\rm L}/D_{\rm L}$ would be minor, so that the numbers for $\Delta w$ would not change very much. Weak lensing will dominate the error on $w$ irrespective of whether the lower cut-off frequency is set to $10^{-5}$ Hz or $10^{-4}$ Hz.

\section{Discussion}

Binary supermassive black hole mergers are ``standard sirens". From the inspiral gravitational wave signal, one can obtain the luminosity distance $D_{\rm L}$ with great accuracy, as well as an estimate of the sky position. If the sky error box is small enough for the host galaxy or cluster of the merger event to be located, then the redshift $z$ can be obtained by electromagnetic means. Using the delicate relationship between $D_{\rm L}$ and $z$, one can then infer the values of the cosmological parameters, in particular $w$. 

As was already shown in \cite{AISSV}, binary SMBH merger events at a redshift $z < 1$ could be used to measure $w$ to within a few percent, provided that one works with the full, amplitude-corrected waveform as opposed to the restricted one, especially if the lower cutoff in LISA's sensitivity is rather sharp. Here we used the most up-to-date waveform model (except for the effect of spin), which is of 3.0PN order in amplitude and 3.5PN in phase, leading to eight harmonics \cite{3PN}\footnote{Recently our own computer code for parameter estimation with amplitude-corrected waveforms was compared with that of two other groups who had independently developed similar code. When taking into account minor differences in waveform approximants, as well as the different choices for LISA's sensitivity curve, excellent agreement was found \cite{LISAPE}.}. The higher signal harmonics in the FWF carry a great deal of additional information as compared to the RWF, so that errors on all parameters are reduced by sizeable factors. In particular, the improvement in the estimation of sky position makes it possible to identify the host in a large fraction of cases \cite{AISSV,SintesAndTrias}. 

What had not been taken into account before is that once the exact sky location has been found in this way, one can repeat the analysis, this time only matched-filtering the data against a family of templates for which sky location is fixed to be that of the host. Since the sky position angles are strongly correlated with luminosity distance, a smaller error on $D_{\rm L}$ is obtained, and a tighter bound can be placed on $w$. In the examples we gave, $\Delta w$ would improve by a factor of 2 or 3, down to less than a percent in some cases. Weak lensing would then be the only limiting factor in measuring $w$, as the error in $D_{\rm L}$ it induces would dominate the uncertainty due to LISA's instrumental noise.

Obviously our investigation is of a preliminary nature. To arrive at an error on $w$ we neglected uncertainties on $H_0$, $\Omega_{\rm M}$, and $\Omega_{\rm DE}$. In principle, all four parameters should be estimated together, with priors from other observations with LISA and ground-based instruments. These could be observations of several different binary SMBH inspiral events, EMRIs \cite{Hogan}, or stellar mass inspirals as seen in Advanced LIGO or Einstein Telescope; this is currently under investigation. Additionally, we made the assumption that $w$ is constant, which need not be the case; here too there is scope for further work. Finally, we only presented a limited number of examples. Large simulations are called for in order to probe the parameter space in a comprehensive manner, the work of Trias and Sintes \cite{SintesAndTrias} could be used for this purpose. That being said, our results already strongly indicate that LISA could be used to measure $w$ with the same accuracy as conventional dark energy probes, but without having to rely on the lower rungs of the cosmic distance ladder.  

\section*{Acknowledgements}

For useful discussions and encouragement we would like to thank 
Bernard Schutz.  We are grateful to the members of the LISA Parameter 
Estimation Task Force, especially Miquel Trias and Alicia Sintes, 
whose comparisons helped validate our {\tt Mathematica} code.
This research was supported in part by STFC grant PP/F001096/1.

\section*{References}
%\begin{the bibliography}
\bibliography{ref-list}
%\end{thebibliography}
\end{document}